\documentclass[12pt]{iopart}
\usepackage[english]{babel}
\usepackage[latin1]{inputenc}
\usepackage{iopams,setstack}
\usepackage{graphicx}
\usepackage{bm}

\begin{document}
\title[Pseudospin Quantum Computation in Semiconductor Nanostructures]
{Pseudospin Quantum Computation in Semiconductor Nanostructures}
\author{V.W. Scarola, K. Park and S. Das Sarma}
\address{Condensed Matter Theory Center, Department of Physics,
  University of Maryland, College Park, MD 20742}

\begin{abstract}
We review the theoretical aspects of pseudospin quantum computation using
vertically coupled quantum dots in the quantum Hall regime.  We
discuss the robustness and addressability of these 
collective, charge-based qubits.  The low energy Hilbert space of a 
coupled set of qubits yields an effective 
quantum Ising model tunable through external gates.  An experimental
prediction of an even-odd effect in the Coulomb blockade spectra of
the coupled quantum dot system allows for a probe of the 
parameter regime necessary for realization of these qubits.    
\end{abstract}


\section{Introduction}
Implementation of useful quantum algorithms requires large scale quantum
information processing.  One perceived advantage of
solid state quantum computing proposals has been the possibility of
scaling up the system to produce nearly homogeneous arrays of 
qubits with tunable interactions.  Some rather promising proposals
\cite{LossDiVincenzo,Kane} make use of real spin in semiconductor 
nanostructures as a
natural two-level system with tunable couplings.  A potential
advantage of real spin quantum computation, over charge-based proposals,
is the long decoherence times for spin states in solids, $~\sim \mu$s
or longer.  One particular demerit of these proposals is the
difficulty in addressing a single spin.  Experimental techniques
required to perform local manipulation of a single
spin through applied magnetic fields push the limits of current
technology \cite{Hu}.  Similarly, single spin detection has
proven difficult.  Recent 
experiments \cite{detection1,detection2,detection3,detection4} 
involving single spin detection have shown some success.  
These measurements are an important first step in
quantum computing with real spins but remain far from the goal of measuring
several individual spins at specific locations.

An interesting solid state quantum computing 
implementation \cite{Yang,Scarola_pseud} makes use of the 
coherent properties of bilayer quantum Hall states confined to
nanostructures offering the possibility of charge-based quantum
computing with long decoherence times.  Charge-based proposals yield
the advantage of addressability.  Detection and manipulation of
individual charges in quantum dots with single electron transistors is
standard practice \cite{SET1,SET2}.  Single charges in 
the solid state, however, are particularly sensitive to 
electric field fluctuations, phonons, and other very strong and common
noise sources in solids.  As we will discuss below, the charge degree
of freedom in quantum Hall states comprise many-body states which map
onto a pseudospin and remain robust 
against external perturbations.  The quantum Hall
liquid is incompressible and, by definition, 
possesses a gap to excited states in the charge sector.  Quantum Hall
states confined to nanostructures \cite{tunelling,MDD,phases} 
therefore offer a many-body charge-based qubit which should 
be less susceptible to certain types of environmental noise than 
similar systems using single charges.  These 
systems are in direct analogy to the Cooper pair box
system \cite{Nakamura1,Nakamura2,Nakamura3} 
(another solid state, charge-based qubit) as a 
mesoscopic, scaled-down version of a coherent, bulk
condensate in the solid state. 

In what follows we review the microscopic theory and fundamental
aspects of pseudospin quantum computing using bilayer quantum Hall
systems confined to nanostructures\cite{Yang,Scarola_pseud,Park}.  In 
Section~\ref{Bilayer} we
discuss the coherent properties of bilayer quantum Hall systems.  We
draw an analogy between exciton condensation in 
these systems and low temperature superconductivity.  In Section~\ref{BQHD} we 
discuss the microscopic theory used to model vertically coupled 
quantum dots in the quantum Hall regime.  We discuss the parameters 
necessary to define a two-level system.  In Section~\ref{qubits} we
construct an effective, single pseudospin model for the qubit.  We
outline several theoretical results which compare noise issues 
of the many-body pseudospin qubits discussed here with similar
single-charge qubits.  In Section~\ref{Coulomb} we review 
a derivation of a low-energy, pseudospin model describing Coulomb
coupled pseudospin qubits.  The resulting quantum Ising model is
sufficient for carrying out a universal set of quantum gates on a
set of pseudospin qubits.  We conclude in Section~\ref{Conclusion}.

\section{Exciton Condensation in Bilayer-Quantum Hall Systems} \label{Bilayer}

In this section we review the physics of exciton condensation in
``bulk'' quantum Hall bilayers.  We discuss the geometry, formalism,
and phenomena in the bulk that will be relevant for our discussion of 
pseudospin quantum computation using a scaled down version of the bulk
system: vertically coupled quantum dots in the quantum Hall regime.
The bulk system consists of two parallel two-dimensional electron gases
separated by a barrier of thickness $\sim 10-100 \AA $, for example.  A 
magnetic field oriented perpendicular to the plane
quantizes the planar motion of electrons into Landau levels (LLs).  Along
the direction perpendicular to the two dimensional plane 
the electrons lie in the lowest sub-band of
the confinement potential.  The finite extent of the wavefunctions in
the perpendicular direction allows for a small amount of single
particle tunneling, $t$, between
the two two-dimensional electron gases. (The tunneling is 
equal to the symmetric-antisymmetric gap 
established by the double quantum well confining the electrons 
perpendicular to the plane.)  At low tunneling the
interlayer physics is dominated by the Coulomb interaction.  Even 
without single particle tunneling the two layers correlate through
the Coulomb interaction.  In fact
the large Coulomb interaction, along with the large magnetic fields in
these systems, polarizes the real electron spin in parameter regimes where
the ground state is essentially ferromagnetic.  In what follows we take
the system to be fully real-spin polarized.    

Transport experiments on quantum Hall bilayers display a variety of
spectacular phenomena  \cite{Perspectives}.  We discuss 
results associated with magnetic fields
large enough to produce one flux quanta per electron, i.e. total
filling $\nu_T=1$.
Interlayer tunneling conductance measurements at this filling 
\cite{Eisenstein1} reveal a dramatic increase 
in tunneling conductance.  The dramatic increase has been 
associated with a spontaneously interlayer  
coherent phase supported by an equal number of electrons and 
correlation holes residing in each of the two dimensional
layers\cite{Perspectives}.  The resulting exciton
condensate has been the subject of intense 
theoretical and experimental study \cite{Perspectives,Eisenstein2}.  

The fundamental properties of this neutral superfluid have been well
established \cite{Perspectives}.  We begin with a Hamiltonian 
describing electrons confined
to bilayer systems:     
\begin{eqnarray}
H=\frac{1}{2m^*}\sum_j\left( i\hbar\bm{\nabla}_j
-\frac{e}{c}\bm{A}_j\right)^2
+V_C
+H_{t}, 
\label{H}
\end{eqnarray}
where the interaction is given by: 
\begin{eqnarray}
V_C=
\frac{e^2}{2\epsilon}\sum_{(i,j);(\alpha,\beta)}^{\prime} 
\frac{1}{\sqrt{\vert \bm{r}^\alpha_{i}-\bm{r}^\beta_{j} \vert ^2
+d^2(1-\delta_{\alpha,\beta})}}.
\end{eqnarray}
Here $m^*$ is the electron effective mass, $\epsilon$ is the
dielectric constant of the host material and $d$ is the
center-to-center interlayer separation.  The indices 
$\alpha,\beta\in\{\uparrow,\downarrow\}$ denote layer index (up or
down) while the prime on the sum indicates $i\neq j$ when $\alpha=\beta$.
We work in the symmetric gauge at magnetic field $B$:
$
\bm{A}=\frac{B}{2}(y,-x).
$ 
$H_t$, in Eq.~\ref{H}, denotes the interlayer tunneling Hamiltonian
which we take to be small.  

The single particle energy spectrum is split into LLs.  The
splitting is given by $\hbar\omega_c$, where
$\omega_c=eB/m^*c$.  At large fields the kinetic energy is quenched to
the lowest LL.  The basis states are given by:   
\begin{eqnarray}
\phi_m=(2\pi 2^m m!)^{-1/2}\left(\frac{z}{l_B}\right)^m 
\exp{(-\vert z \vert^2 /4l_B^2)},
\label{basis}
\end{eqnarray}
where the planar coordinates $z=x+iy$ scale with the magnetic length 
$
l_B=(\hbar c/eB)^{1/2}.
$
The quantum number $m$ represents the angular momentum.  It is 
the eigenvalue of the angular momentum operator:
$
\hat{L}_z= z\partial_z-z^*\partial_{z^*}.
$
Recast in the basis of the lowest LL the problem simplifies.  
Estimates \cite{Price} of the effects of 
LL mixing (along with finite thickness perpendicular 
to the plane) find, at most, a 15\% correction to energy
differences.  In what follows we ignore finite thickness and LL
mixing. 

At $\nu_T=1$ the Hartree-Fock solution of Eq.\ref{H} provides a
surprisingly accurate solution over a 
large range of parameters, $d/l_B\lesssim 1.5$.  The layer 
index enlarges the Hilbert space.
A Hund's rule, applicable to layer index, picks out a single low
energy state, the ground state in the Hartree-Fock approximation, to
minimize the Coulomb energy cost.  The 
total ground state wavefunction, including the orbital and layer
degrees of freedom, is generally given by: 
$
\Psi={ \cal A}\left[ \prod_{\{ \alpha \},\{ m\}} \phi_m \chi_{\alpha}\right].
$
Here 
${ \cal A}$ is the antisymmetrization operator and $\chi$ is a spinor
dependent on the set of all layer indices $\{\alpha\}$.  The orbital part  
takes a simple form \cite{Halperin}:
\begin{eqnarray}
\psi_{S_z}\sim\prod_{r,s}(z^{\uparrow}_{r}-z^{\downarrow}_{s})
\prod_{i<j} (z^{\uparrow}_i-z^{\uparrow}_j)
\prod_{i'<j'}(z^{\downarrow}_{i'}-z^{\downarrow}_{j'}) 
\exp \left(-\sum_{i,\alpha} \vert z_i^{\alpha} \vert^2 /4l_B^2
\right).
\label{111}
\end{eqnarray}
This solution represents the exact orbital part of the ground state in the limit
$B\rightarrow\infty$ and $d/l_B\rightarrow 0$.  This limit ensures our lowest LL projection
while effectively lowering the interlayer separation in units of $l_B$.
From this form of the wavefunction we see that each electron lies opposite a
correlation hole in the neighboring layer.  The electron and its
opposing zero form a neutral electron-hole pair.  These excitons condense to
from an exciton condensate associated with a spontaneously broken
symmetry, discussed below.   

The spinors in the total wavefunction suggest a pseudospin
interpretation of the layer index.  We formally define the pseudospin
to be:  
\begin{eqnarray}
\hat{\bm{S}}=\frac{1}{2}
\sum_{m} c^{\dagger}_{m,\alpha} \bm{\sigma}_{\alpha\beta} c^{\vphantom\dagger}_{m,\beta},
\end{eqnarray}
where $c^{\dagger}_{m,\alpha}$ creates an electron in layer $\alpha$
with orbital angular momentum $m$ and $\bm{\sigma}$ are the usual
Pauli matrices.  With this definition the eigenvalue of the pseudospin
operator along the pseudospin $z$ direction, 
$S_z=(N_\uparrow-N_\downarrow)/2$,  
denotes the relative number difference between layers.  Along the
pseudospin $x$ direction the pseudospin operator is 
equivalent to the interlayer
tunneling operator.  $\hat{S}_x$ creates a bonding state between
layers.  The pseudospin operator along the pseudospin $y$ 
direction, $\hat{S}_y$, is equivalent to an interlayer current operator.   

Consider a bilayer system with an interlayer bias adjusted to ensure
an equal number of electrons in each layer, on average.  The many-body
ground state will consist of a coherent superposition of orbital states
$\psi_{S_z}$ centered around $S_z=0$.  The electrons and
zeroes swap places in a coherent fashion, enhancing the
interlayer tunneling in the process.  
The coherent properties of the ground state can be seen explicitly in
it second quantized form:
\begin{eqnarray}
\prod_{m}\left[ 
1+c^{\dagger}_{m,\uparrow}c^{\vphantom\dagger}_{m,\downarrow}\right] 
\prod_{m'} c^{\dagger}_{m',\uparrow}\vert 0\rangle.
\label{exciton}
\end{eqnarray}
This ground state has a form similar to the well known
Bardeen-Cooper-Schrieffer (BCS)
ground state of a superconductor: 
\begin{eqnarray}
\prod_{k}\left[ 
u_k+v_k
c^{\dagger}_{k,\uparrow}c^{\dagger}_{-k,\downarrow}\right] 
\vert 0\rangle,
\label{BCS}
\end{eqnarray}
where $k$ denotes wave-vector, the arrows indicate real 
spin (in Eq.~\ref{BCS}), and the
coefficients $u$ and $v$ are fixed to ensure proper relative 
phases.  By comparison we see that, with a redefined vacuum 
($\prod_{m'} 
c^{\dagger}_{m',\uparrow}\vert 0\rangle \rightarrow \vert 0\rangle $), the excitons
condense to form a neutral superfluid analogous to a 
condensation of cooper pairs in a BCS
superconductor.  Eq.~\ref{exciton} describes a coherent superposition
of eigenstates of pseudospin centered around $S_z=0$ while the BCS
state captures a coherent superposition of number eigenstates.  

Exciton condensation is
associated with a spontaneous breaking of pseudospin symmetry.  The
electrons must have some (albeit arbitrarily small)
interlayer, single particle tunneling to enable passage between layers.
But even with an infinitesimally small amount of 
tunneling, $t/(e^2/\epsilon l_B) \rightarrow 0$, the system, 
in the coherent ground state, exhibits a large 
tunneling renormalized by the interaction:
\begin{eqnarray}
\lim_{t\rightarrow0}\langle \Psi \vert \hat{S}_x \vert
\Psi\rangle\sim N/2, 
\label{order}
\end{eqnarray}
where $N$ is the total number of particles.  
An arbitrarily small amount of tunneling spontaneously reorients the
total pseudospin along the pseudospin $x$ direction.  
The observation \cite{Eisenstein1} of enhanced 
interlayer conductance, at $\nu_T=1$, suggests the formation
of an exciton condensate with an order parameter defined by 
Eq.~\ref{order}.  

\begin{figure}
   \centering
   \includegraphics[clip,totalheight=5truecm]{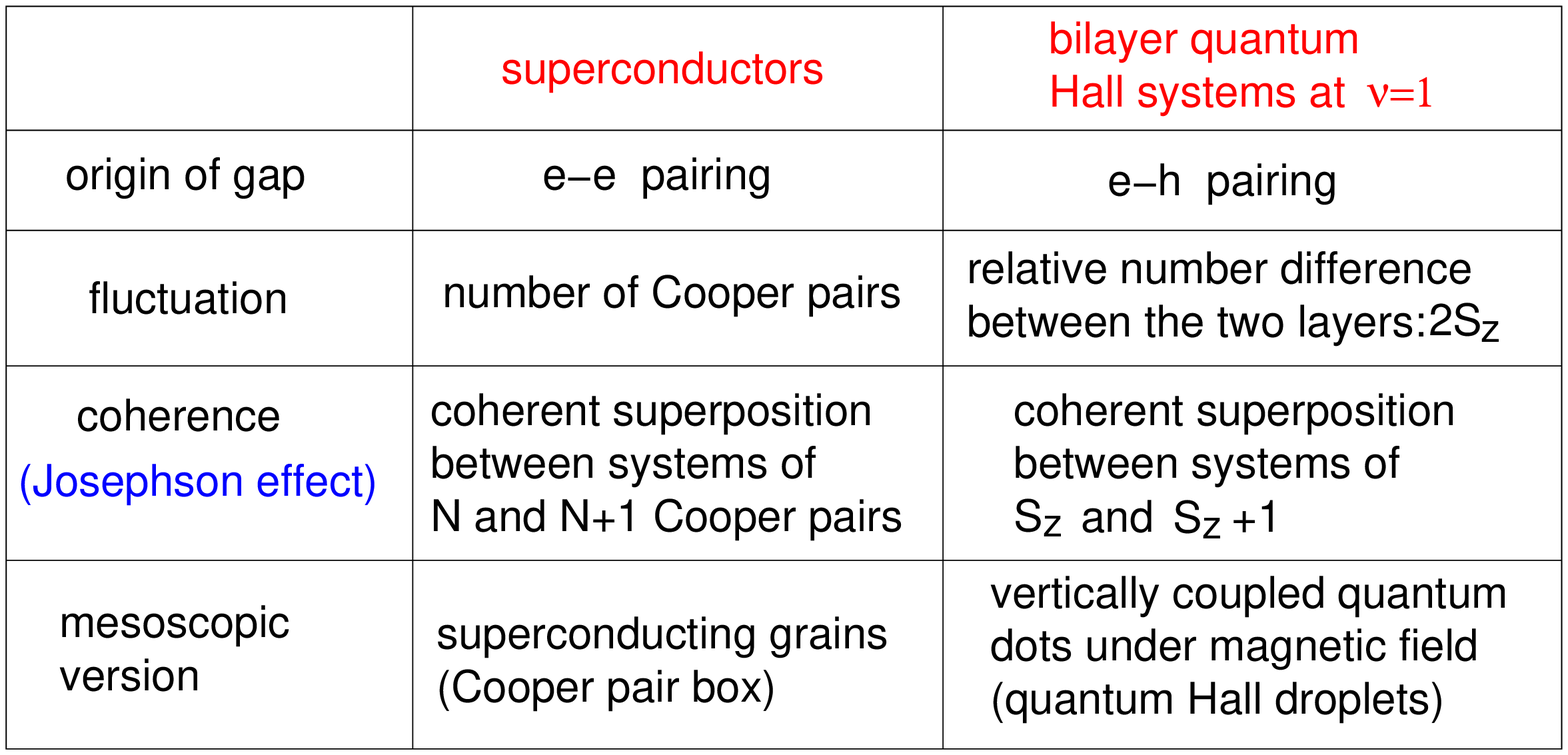}  
   \caption{Table summarizing the analogous properties of
   superconductors and exciton condensation in bilayer quantum Hall
   systems.} 
  \label{analogy}
\end{figure}

By analogy with superconductivity one may consider the mesoscopic
version of the exciton condensate.  The mesoscopic version of a BCS
superconductor is a superconducting grain which exhibits some of
the coherent properties of superconductors but in a smaller system.
Vertically separated, lateral quantum dots in the quantum Hall regime
offer the ``mesoscopic'' version of the bulk exciton condensate
discussed above.  Figure~\ref{analogy} summarizes the analogous
properties of exciton condensates and superconductors. 
In what follows we discuss the possibility of encoding quantum
information in the layer degree of freedom of a mesoscopic exciton
condensate.  In the next section we discuss the 
properties of an individual qubit and 
the experimental prediction of an even-odd effect in the Coulomb
blockade spectra caused by spontaneous interlayer phase 
coherence inherent in our qubit.   

\section{Bilayer Quantum Hall Droplets}
\label{BQHD}

We now define the model and parameter regime necessary to establish a
robust two-level system in a mesoscopic version of the exciton
condensate discussed above.  We review a proposal \cite{Park} 
designed to characterize the two-level system and quantitatively test the 
interlayer coherence present in the system.  An
even-odd effect in the Coulomb blockade spectra of such a device
yields an accurate test of interlayer coherence and, as we will see,
provides important information related to the two-level systems encoded in the 
pseudospin degree of freedom. 
  
Consider a bilayer quantum Hall system at $\nu_T=1$ with vanishingly
small interlayer tunneling, as described in
the previous section.  The application of an external confinement 
fixes the total number of particles 
$N=N_{\uparrow}+N_{\downarrow}$.  It is sufficient
to consider parabolic confinement:
\begin{eqnarray}
H_{\omega_0}=\frac{1}{2}m^*\omega_0^2\sum_{i}\vert \bm{r}_i\vert^2,
\end{eqnarray}          
where the confinement parameter, $\hbar\omega_0=3-8$meV in GaAs samples, can be
adjusted with external gates.  In the presence of a parabolic
confinement the single particle eigenstates of the 
non-interacting Hamiltonian are the Fock-Darwin
states \cite{Fock,Darwin}.  In the large magnetic field limit, 
$\omega_c\gg\omega_0$, the
eigenstates reduce to Eq.~\ref{basis} but with the replacement:
$l_B\rightarrow a$, where we define the modified magnetic length,  
$
a\equiv l_B (1+4\omega_0^2/\omega_c^2)^{-1/4}.
$
These states provide a basis for our calculations.  

With a small number of particles $N\sim 1-100$ the system forms a 
maximum density liquid with an equal number of particles and 
flux quanta piercing the system.  This few-body state, often 
called the maximum density
droplet (MDD), is separated from excited states by a gap.  It is the
mesoscopic realization of the bulk $\nu_T=1$ quantum Hall state.  In the
special case of an odd number of particles the two, degenerate 
low energy states have total pseudospin $S_z=\pm 1/2$.  The $+$ ($-$)
indicates one excess charge in the top (bottom) layer.  These two
states, denoted $\vert \pm 1/2 \rangle$, define a two-level system
arising from a competition between the Coulomb interaction and
confinement.  An even number of particles distributed between the two
layers yields one state with $S_z=0$.               

\begin{figure}
   \centering
   \includegraphics[clip,totalheight=8truecm]{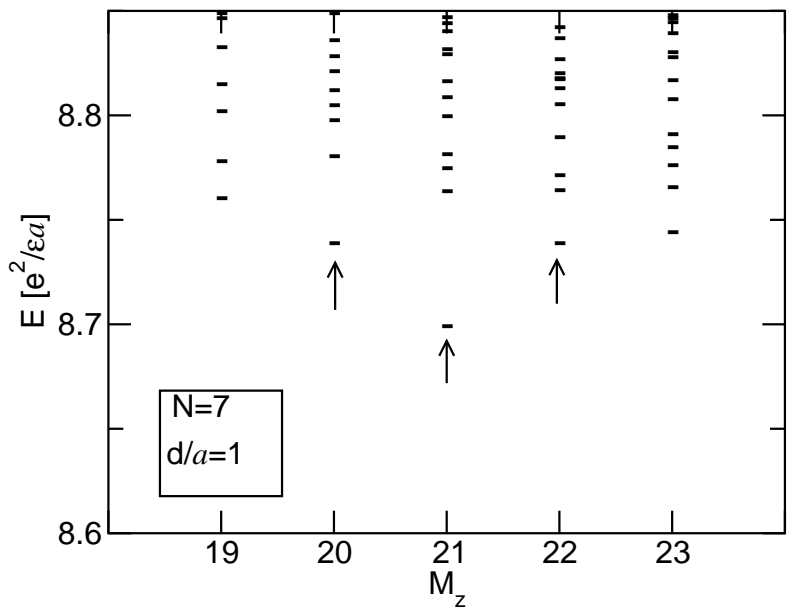}  
   \caption{Energy spectrum as a function of total angular momentum
   for a bilayer quantum Hall droplet with four (three) electrons in
   the top (bottom) layer, $S_z=+1/2$.  The energies are obtained from
   exact diagonalization of Eq.~\ref{H_second} with tunneling, $t=0$,
   interlayer separation, $d=a$, and 
   $\gamma=0.1187 e^2/\epsilon a$.  We define the 
   parameter $\gamma \equiv \hbar\left( \sqrt{\omega^2_c +4\omega^2_0}
   -\omega_c \right)/2$ and the modified magnetic length $a\equiv l_B
   (1+4\omega_0^2/\omega_c^2)^{-1/4}$ in terms of the confinement parameter,
   $\omega_0$, the cyclotron frequency, $\omega_c$, and the magnetic
   length $l_B=(\hbar c/eB)^{1/2}$.  The 
   arrows indicate the three
   lowest energy states.  The central arrow indicates the 
   $\vert +1/2 \rangle$ state of our qubit. 
   (From Ref.~\cite{Park})}
  \label{spectrum}
\end{figure}

We study the low energy physics of a bilayer quantum Hall
droplet (BQHD) with an odd number of particles using exact
diagonalization.  In second quantization the lowest LL Hamiltonian 
describing the system, including confinement and finite 
tunneling now reads:
\begin{eqnarray}
H_{LLL}=\gamma\sum_{m,\alpha}mc^{\dagger}_{m,\alpha}c^{\vphantom\dagger}_{m,\alpha}
+\sum_{\{m\},(\alpha,\beta)}V_{\{m\}}^{\alpha,\beta}
c^{\dagger}_{m_1,\alpha}c^{\dagger}_{m_2,\beta}
c^{\vphantom\dagger}_{m_3,\beta}
c^{\vphantom\dagger}_{m_4,\alpha}
-t\hat{S}_x,
\label{H_second}
\end{eqnarray} 
where the confinement redefines the basis length scale and adds the first term
with coefficient 
$
\gamma \equiv \hbar\left( \sqrt{\omega^2_c +4\omega^2_0}
-\omega_c \right)/2.
$
Explicit expressions for the Coulomb matrix elements,
$V_{\{m\}}^{\alpha,\beta}$, can be found in the literature
\cite{Hawrylak,Wojs}.  We diagonalize the system in the limit $t=0$
and consider finite tunneling effects subsequently.  In this limit the
total $z$ component of pseudospin, $S_z$, the total angular momentum,
$M_z$, and the total particle number are all good quantum numbers.
The first two terms in Eq.~\ref{H_second} compete.  It is
necessary to tune the magnetic field and confinement to the proper 
values ensuring that the MDD is the ground state of the
system.  The angular momentum of the MDD (the orbital state defined by
Eq.~\ref{111}) is $M_z=N(N-1)/2$.  With $N=7$, for example, we find
that a range of confinements centered around $\gamma=0.1187
e^2/\epsilon a$ yield the MDD ground state.  Fig.~\ref{spectrum} 
plots the low energy Hilbert space of Eq.~\ref{H_second} for $N=7$ with 
$t=0$ in the $S_z= + 1/2$ sector as a function of total angular 
momentum.  This implies four (three) electrons in the top (bottom) layer.  
The interlayer separation is chosen to be $d=a$.    
The arrows indicate the three lowest energy states.  The central arrow 
shows the $\vert +1/2 \rangle$ ground state.  
The left and right-most
arrows indicate the edge excitations.  
It was
found in Ref.~\cite{Park} that the gap to edge 
excitations remains finite for $d/a\lesssim 2.0$.  The many-body
ground state found here is one of two degenerate states which form a
two-level system, separated from excited states (with the same $S_z$)
by $\sim 0.07e^2/\epsilon a$ at $d=a$.  

For an odd number of particles, BQHD states with $S_z=\pm
1/2$ have the lowest charging energy cost.  
They are separated from states with higher $\vert S_z \vert$ by 
the relative charging energy cost: $\alpha \hat{S}_z^2/N$, where we
find \cite{Park} $\alpha/(e^2/\epsilon a) \simeq -0.18 + 0.35 d/a$ 
for $d/a \gtrsim 0.5$ through an empirical fit to our numerical, exact
diagonalization over several values of $N$.  We find that for $d\approx
a$ the two-level system defined by $\vert \pm 1/2\rangle$ remains
separated from excited states by an energy $\sim 0.05 e^2/\epsilon a $.

\begin{figure}
   \centering
   \includegraphics[clip,totalheight=6truecm]{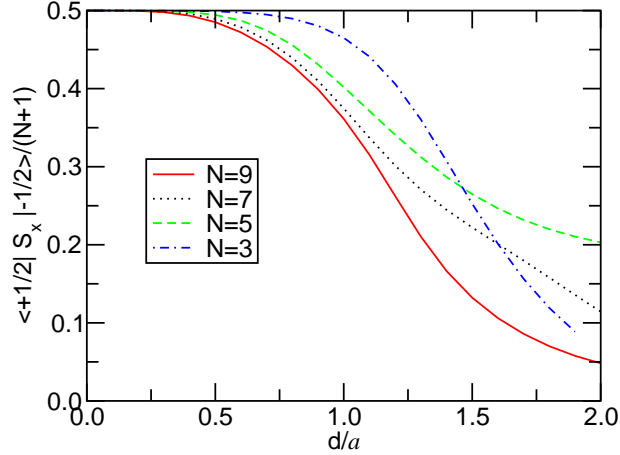}  
   \caption{The interlayer coherence measure, as defined 
     in Eq.~\ref{coherence}, plotted as a function of
     inter-layer distance for several particle numbers with zero tunneling.  } 
  \label{sx}
\end{figure}

We now discuss the coherence properties of a BQHD in the MDD state.
As for the bulk system, the BQHD will develop a renormalized
tunneling gap: $\Delta_x = t \langle +1/2 | \hat{S}_x | -1/2
\rangle$, in the limit  $t/(e^2/\epsilon a) \rightarrow 0$.  The
Coulomb interaction enhances the 
single particle tunneling by a factor of order $N$.  The 
definition of spontaneous interlayer phase coherence in 
the BQHD is then:
\begin{eqnarray}
\lim_{t\rightarrow 0} \frac{\Delta_x}{t} =
\lim_{t\rightarrow 0} \langle +1/2 | \hat{S}_x | -1/2 \rangle
\neq 0.
\label{coherence}
\end{eqnarray}
In Fig.~\ref{sx} we plot $\Delta_x/t$ as a function of interlayer
separation for several odd particle numbers in the MDD state.    
We find  $\Delta_x$ to be sizable for $d/a \lesssim 1.0$.  The quantity 
$\Delta_x$ provides a mesoscopic measure of spontaneous interlayer
phase coherence with direct experimental relevance.  

\begin{figure}
   \centering
   \includegraphics[clip,totalheight=9truecm]{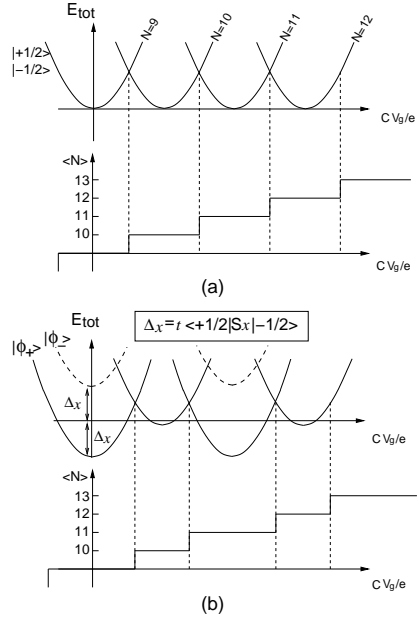}  
   \caption{Schematic depicting the even-odd effect in bilayer quantum
   Hall droplets.  The total energy is plotted as a function of $C
   V_g/e$, where $V_g$ is the gate voltage and $C$ the lead-dot
   capacitance.  The total energy includes the charging energy.  The
   average number of electrons distributed between the two dots is
   denoted $\langle N \rangle$.  The top panel, (a), depicts the
   situation where there is no interlayer coherence.  The states
   $\vert \pm
   1/2\rangle$ are degenerate here.  The bottom panel, (b), depicts
   the even-odd effect due to a coherence gap, $\Delta_x$.  The
   degenerate set of states here splits into symmetric and
   antisymmetric 
   combinations $\vert \phi_+\rangle$ and $\vert \phi_-\rangle$.} 
  \label{evenodd}
\end{figure}

The coherence order parameter, $\Delta_x$, may be related 
to the Coulomb-blockade peak spacing measured 
in experiments on BQHDs.  Coulomb-blockade 
peaks in the tunneling conductance of quantum dots
arise when the gate voltage, $V_g$, is tuned so that the total
energy of the $N$ electron system coincides with the energy of the
$N+1$ electron system.  The total energy of the 
bilayer quantum dot system includes the total 
charging energy cost:
\begin{eqnarray}
H_{C} = \frac{e^2}{2C_{\Sigma}} \left( N - \frac{C V_g}{e} \right)^2,
\end{eqnarray}
where $C_{\Sigma}$ is the total capacitance of the double dot system and
$C$ is the lead capacitance to one of the two dots.  
The enhanced tunneling gap leads to a splitting between hybridized
states formed from the states
$\vert \pm 1/2\rangle$ in systems with $N$ odd, thereby modifying the
total energy of the system (see Fig.~\ref{evenodd}).  For an even number
of particles no such splitting exists.  There is only one state which
minimizes the charging energy cost giving: 
$\langle 0 \vert \hat{S}_x\vert 0 \rangle =0$.  Therefore the
Coulomb-blockade peak spacing will alternate with an even or odd
number of particles.  Fig.~\ref{evenodd} illustrates the even-odd
effect in BQHDs.
The top panel, (a), shows the situation with no
interlayer coherence: $d/a\gg 1$.  The bottom panel, (b), 
illustrates the even-odd effect arising from finite $\Delta_x$.
Observation of an even-odd effect in the Coulomb blockade peak spacing
would provide evidence for conditions sufficient to establish 
a two-level system in a BQHD.

\section{Pseudospin Qubits}
\label{qubits}

The many-body two-level system defined by the states $\vert
S_z=\pm1/2\rangle$ in a BQHD offers a charge-based qubit with tunable 
{\em effective} magnetic fields.  In this section we discuss two important
aspects of BQHD qubits: the tunability and robustness of the qubit.
The qubit proposal outlined here has several advantages and
disadvantages in meeting the DiVincenzo criteria \cite{DiVincenzo}, as 
opposed to other solid state quantum computing proposals.  The
disadvantages include susceptibility to leakage (as compared to real
spin \cite{LossDiVincenzo} which is inherently a two-level system) and 
charge noise.  The advantages include ease of 
addressability (detection and manipulation of a single electron
charge, as opposed to a single electron spin), tunability 
thorough externally applied electric fields, and a certain
rigidity against external perturbations (as compared to 
single-electron, charge-based qubits).  

In the reduced Hilbert space $\vert \pm 1/2 \rangle$ (as defined for a
BQHD with an odd number of particles in the MDD state) two parameters 
may rotate the pseudospin.  As noted previously: finite 
interlayer tunneling effectively rotates the pseudospin 
along the pseudospin $x$ direction.  The parameter $t$ can be 
tuned with external
gates or by application of a real, in-plane magnetic field.  Both
fields alter the spread of the electron density along the direction
perpendicular to the two-dimensional plane thereby tuning the
interlayer overlap.  These fields provide an effective magnetic field
along the pseudospin $x$ direction which is always negative.

An external bias, $\Delta_z$, applied perpendicular to the two 
dimensional plane
of a BQHD energetically favors one pseudospin state over another.
$\Delta_z$ acts as an effective magnetic field along the pseudospin
$z$ direction.  We may therefore rotate the pseudospin direction
through any point on the Bloch sphere by pulsing $\Delta_z$ 
and $\Delta_x$.  In the reduced Hilbert space, $\vert \pm 1/2 \rangle$, 
we have a reduced Hamiltonian \cite{Yang,Park}:
$
H_{R} = - \Delta_x \sigma_x +\Delta_z \sigma_z.
$
The collective state associated with an excess charge in either the
top or bottom layer of a coherent BQHD in the MDD state therefore
provides a fully tunable qubit.  

Several potential error sources relevant for a BQHD qubit have been
quantitatively addressed in the literature \cite{Yang, Scarola_pseud}.  
Three types of error sources were studied 
in Refs.~\cite{Yang, Scarola_pseud}: I) spatially local 
decoherence  II) spatially
global decoherence III) leakage (as opposed to decoherence) 
due to density perturbations.  The first 
type was studied in Ref.~\cite{Yang}.  It was shown 
that phase flip errors arising from 
an inhomogeneous, externally applied potential are
strongly suppressed by increasing the total number of particles
defining the BQHD qubit.  The results shows that phase flip errors are
suppressed by more than an order of magnitude by increasing $N$ from 1
to 10.

The second type of error was studied in Ref.~\cite{Scarola_pseud}.   
The analysis of phase flip errors arising from fluctuations in
spatially homogeneous, external voltages in the 
leads (fluctuations in $\Delta_z$) finds that, at low 
temperatures, the decoherence rate (the ratio of the dephasing rate to
the elementary logic operation rate) can be made small enough to allow
for fault-tolerant quantum computation.  More precisely, we find 
the ratio to be:  
$  
4\left[C/(2C+C_\Sigma)\right]^2 R_v/R_K,   
$
where $R_v\sim 50 \Omega$ is the
typical impedance of the voltage circuit and $R_K=h/e^2$.  The above
analysis applies to voltage noise in leads attached to any charge
based qubit, pseudospin or otherwise, modeled in 
the spin-boson formalism.  The above
formula captures an implicit fact; isolating the system from 
the environment, $C\ll C_\Sigma$, can reduce the decoherence rate.  

In addition to dephasing from voltage fluctuations, there is also the
possibility of leakage due to density perturbations such as phonons.
In Ref.~\cite{Scarola_pseud} it was shown that increasing the particle
number suppresses the form factor associated with phonons 
coupling the states $\vert \pm 1/2 \rangle$ to unwanted edge modes,
$\vert e \rangle$.
The form factor: 
\begin{eqnarray}
\vert \langle e \vert \sum_{j=1}^N \exp(i \bm{k}\cdot \bm{r}_j) \vert 
\pm 1/2 \rangle \vert^2,
\end{eqnarray}
was suppressed by more than an order of magnitude
in increasing $N$ from 1 to 9.  Here $\bm{k}$ is the wave-vector of a
density perturbation in the plane caused, for example, by a phonon of
wave-vector $\bm{k}$.  We emphasize that the diminished form factor 
suppresses leakage due to {\em any}
density perturbation, including both acoustic and optical 
phonons.  The results of Refs.~\cite{Yang, Scarola_pseud} suggest 
that the rigidity inherent in collective
and incompressible BQHD states suppress decoherence and leakage 
mechanisms otherwise
present in single-electron, charge-based qubits in quantum dots.

\section{Coulomb Coupling}
\label{Coulomb}

We now discuss the possibility of constructing a 
universal set of quantum gates by coupling BQHD qubits.  
There are several types of inter-BQHD couplings possible.
Three examples are exchange, capacitive, and Coulomb coupling.  In
what follows we discuss Coulomb coupled BQHDs.   

\begin{figure}
   \centering
   \includegraphics[clip,totalheight=5truecm]{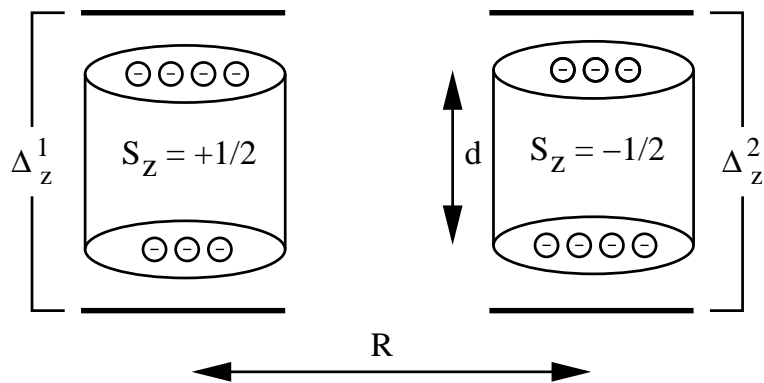}  
   \caption{Schematic illustrating the low energy state of Coulomb
   coupled, neighboring bilayer quantum Hall droplets separated by a
   center-to-center distance $R$.  The
   configuration depicted here represents the pseudospin product state $\vert
   +1/2,-1/2\rangle$.  The quantities $\Delta_z^i$ depict the applied
   bias on the $i$th set of quantum dots.} 
  \label{schematic}
\end{figure}

\begin{figure}
   \centering
   \includegraphics[clip,totalheight=5truecm]{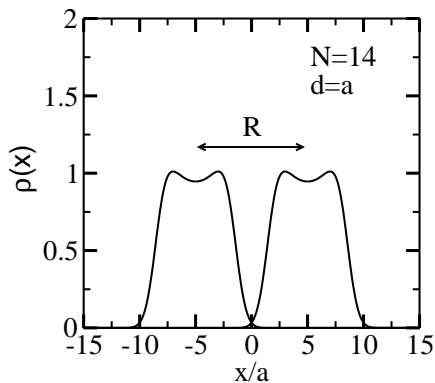} 
   \caption{Normalized density profile versus lateral separation along 
     the $x$-axis for a pair of bilayer quantum Hall droplets with a 
     lateral separation $R=10a$ in the maximum density droplet state, 
     where $a$ is a modified magnetic length defined in the text.  
     The vertical inter-dot spacing is $d=a$.  There are a total of fourteen 
     electrons distributed among all four droplets. } 
  \label{density}
\end{figure}

Consider two neighboring BQHDs, as shown schematically in 
Fig.~\ref{schematic}.  The centers of the BQHDs are placed a distance
$R$ apart. 
The Coulomb interaction between neighboring droplets will
favor an anti-alignment of charge as depicted in 
Fig.~\ref{schematic}.  The charge distributes among 
two parallel disks with an excess of charge in the top (bottom) layer
in the state  $\vert +1/2\rangle$ $(\vert -1/2\rangle)$.  The disk
extends to roughly the largest orbital 
state, $2\sqrt{N} a$.  Fig.~\ref{density} plots the density of two
neighboring BQHDs  placed with a center-to-center distance
of $R=10a$.  We obtain the densities from exact diagonalization of
Eq.~\ref{H_second} for a single BQHD with $t=0$, $d=a$ and $N=7$.

We consider the low energy Hilbert space of two, well separated BQHDs.  
The large inter-BQHD separation ensures little overlap between
electronic states in neighboring BQHDs.  We also require that $R$ is large
enough so that a neighboring BQHD does not induce unwanted intra-BQHD
excitations.  In this limit the Hilbert space of two neighboring 
BQHDs comprises a set of four product states $\{\vert \pm 1/2, \pm
1/2\rangle, \vert \pm 1/2, \mp  1/2\rangle \}$.  In this basis the
resulting inter-BQHD Coulomb interaction maps onto a 
pseudospin Ising interaction:        
\begin{eqnarray}
H_I=\frac{J}{2}\sigma^1_z\sigma^2_z
\end{eqnarray}
where we define the effective exchange splitting to be: 
\begin{eqnarray}
J&=&\langle +1/2,+1/2\mid V(R,d;\bm{r}_i^{\alpha},
\bm{r}_j^{\prime \beta})\mid +1/2,+1/2\rangle
\nonumber
\\ 
&-&\langle +1/2,-1/2\mid V(R,d;\bm{r}_i^{\alpha},
\bm{r}_j^{\prime \beta})\mid +1/2,-1/2\rangle,
\end{eqnarray}
where:
\begin{eqnarray}
&V&(R,d;\bm{r}_i^{\alpha}, \bm{r}_j^{\prime \beta})=
\nonumber
\\
&&\sum_{i,j}\frac{e^2}{\epsilon ' a}
\frac{1}{\sqrt{(x_i^{\alpha}-x_j^{\prime \beta}+R/a)^2
+(y_i^{\alpha}-y_j^{\prime \beta})^2+(d/a)^2(1-\delta_{\alpha,\beta})}}.
\end{eqnarray}
$\bm{r}$ $(\bm{r}^{\prime})$ indicates the radial vector in the $x$-$y$
plane in the left (right) BQHD and $\epsilon '$ is the
inter-BQHD dielectric constant. 
At large distances, $R> 25a$, it was shown \cite{Scarola_pseud} 
that the ``exchange'' coefficient exhibits dipolar
behavior, $J\propto R^{-3}$.  However, at these large distances we
find $J$ to be negligibly small in GaAs devices, 
leaving only nearest neighbor interactions which can be 
tuned with $\epsilon^{\prime}$ or $R/a$.  Nonetheless, the long-range
part of a dipolar coupling (albeit small)
can give rise to errors in quantum gates
constructed entirely from nearest neighbor interactions.  In
Ref.~\cite{DeSousa} it was shown that errors due to dipolar
coupling can be treated with quantum error correction gating schemes.   

We generalize the double BQHD system to an arbitrary number of BQHDs.
Each BQHD can be thought of as a lattice site $i$ containing a pseudospin
interacting with its neighbor through an Ising interaction: 
\begin{eqnarray}
H_{I} = \sum_{i}[ -\Delta_x^i \sigma_x^i +\Delta_z^i \sigma_z^i]
+ \frac{1}{2}\sum_{i,j}J_{ij} \sigma_z^i \sigma_z^j.
\end{eqnarray}
The above model, with tunable coefficients, offers a universal set of
quantum gates \cite{Barenco}.  The effective magnetic fields are tunable 
through external electric fields.  $J$ is not so easily tuned.  But, an
intermediate, idle BQHD with an odd or even number of electrons can
turn the interaction between next nearest neighbors on or 
off.  Another possibility borrows techniques from nuclear magnetic
resonance theory.  It is well known that similar models with 
fixed inter-spin interactions allow quantum
computation with refocusing pulses \cite{NMR1,NMR2}. 

\section{Conclusion}\label{Conclusion}
 
We have reviewed the theoretical aspects of pseudospin quantum
computation using confined, bilayer quantum Hall states.  We argue that
bilayer quantum Hall droplets offer the quantum Hall analogue of
Cooper pair boxes.  The even-odd effect in Coulomb blockade spectra 
provides criteria sufficient in defining a two-level 
system in the low energy Hilbert space
of a bilayer quantum Hall droplet.  The many-body, 
two-level system encoded in the
which-layer degree of freedom provides advantages over single-charge
based qubits when considering external perturbations due to the 
environment.  Coulomb coupling offers the possibility of a universal
set of quantum gates through a pseudospin, quantum Ising model.  

We would like to thank J.K. Jain for valuable discussions.   We
acknowledge support from ARDA and NSA-LPS.

\end{document}